\documentclass[fleqn,usenatbib]{mnras}

\usepackage{newtxtext,newtxmath}

\usepackage[T1]{fontenc}

\DeclareRobustCommand{\VAN}[3]{#2}
\let\VANthebibliography\thebibliography
\def\thebibliography{\DeclareRobustCommand{\VAN}[3]{##3}\VANthebibliography}

\usepackage{graphicx}	
\usepackage{amsmath}	
\usepackage[normalem]{ulem} 

\title[Kilonova impact on GRB VHE emission]{The impact of kilonova seed photons on GRB VHE emission}

\author[John P. Hope et al.]{
John P. Hope,$^{1}$\thanks{E-mail: jph58@bath.ac.uk}
Hendrik J. van Eerten,$^{1}$
\\
$^{1}$Department of Physics, University of Bath, Claverton Down, Bath, BA2 7AY, UK \\
}

\date{Accepted XXX. Received YYY; in original form ZZZ}

\pubyear{\the\year{}}

\begin{document}
\label{firstpage}
\pagerange{\pageref{firstpage}--\pageref{lastpage}}
\maketitle

\begin{abstract}
Over the last several years, there have been a number of GRBs with very high energy (VHE) emission in excess of 100 GeV, and even above 1 TeV, detected. In several instances, synchrotron seed photons do not fully explain the emission observed, suggesting the presence of other seed photon sources to up-scatter. In this work, we consider the kilonova as a source of seed photons for up-scattering in the afterglow. We model the kilonova as a thermal source injecting into the back of a GRB fireball, evolved using a shell model, and with the electron and photon populations updated via a kinetic solver. We find that VHE emission from weaker afterglows, such as those found in short GRBs, can be affected by such seed photons, with the kilonova seed photons mitigating the loss of synchrotron photons on the VHE emission when afterglow parameters are varied. We also find that VHE emission in structured jets, due to weaker synchrotron emission at their wings, can also benefit from this supply of seed photons, especially when viewed off-axis. We apply this model to GRB 170817A, and show that its VHE spectral flux is higher than expected in previous models for the first $100$ days, though still below the detection threshold.
\end{abstract}




\section{Introduction}
Several recent gamma-ray bursts (GRBs)\footnote{GRBs 160821B, 180720B, 190829A, 201015C, 201216C and 221009A.} have been detected with very high energy (VHE) emission in their afterglow (for recent reviews, see e.g. \citealt{universe7120503, TeVReviewTh, TeVReviewObs}). For instance, GRB 180720B had VHE emission detected by the H.E.S.S observatory at 100 -- 440 GeV around $10$--$12$ hours after first detection \citep{2019Natur.575..464A}, the first case of confirmed TeV emission was with GRB 190114C \citep{2019GCN.23701....1M, Ajello_2020}, with GeV and TeV emission observed in the first $20$ minutes of detection \citep{2019Natur.575..455M}, and GRB 221009A, the brightest GRB ever detected \citep{lesage2023fermi, an2023insighthxmt, Frederiks_2023}, was detected by the LHAASO observatory with photons of 3 TeV and above from $230$ -- $900$s post detection, including some in excess of $10$ TeV \citep{doi:10.1126/sciadv.adj2778}. There may be other events that have not been accounted for, though due to various factors \citep{10.1063/1.4968980, Longo:2021wG, TeVReviewObs}, such as poor observing conditions, timing of events and the rarity of the intrinsic properties associated with detectable VHE emission, they have not been seen. Recent improvements to telescope capabilities and newer observatories (such as LHAASO, \citealt{2019arXiv190502773C}) may account for the increased number found over the last few years.

The simplest explanation for VHE emission is that it arises from inverse Compton (IC) up-scattering of the synchrotron photons emitted by shock accelerated electrons, called synchrotron-self Compton (SSC) scattering in the literature. For example, VHE emission in the cases of GRB 190114C \citep{2019Natur.575..455M} and GRB 201216C \citep{10.1093/mnras/stad2958} has been shown to be consistent with SSC. However, in certain cases SSC cannot fully explain the observed emission. In the case of GRB 190829A, the TeV spectrum was flatter than predicted by SSC, leading to proposed mechanisms such as direct synchrotron emission into the VHE regime \citep{doi:10.1126/science.abe8560} and photohadronic interactions between the Fermi accelerated protons and seed photons \citep{Sahu_2022}, which \citet{2020ApJ...895L..41S} have argued can also be applied to GRB 190114C and GRB 180720B.

Another proposed mechanism for generating VHE emission is through external Compton (EC) up-scattering. This term represents any process where electrons up-scatter seed photons which were generated by a source external to the electrons in the GRB blast-wave. Examples of external sources include the internal dissipation/late prompt emission from the central engine (see for example, \citealt{Beloborodov_2005, Murase_2011, Zhang_160821B}), the reverse and/or forward shock with respect to each other \citep{Panaitescu_1998, Wang_2001} or from a cocoon \citep{Kimura_2019} or AGN disk \citep{Yuan_2022}.

In the case of short GRBs (sGRB), and in certain long GRBs (see \citealt{Troja211211A, 2022Natur.612..223R, 2023NatAs...7..976L, Dichiara_2023}) the progenitor neutron star merger can produce a kilonova (KN), a luminous, (semi-)isotropic thermal emission arising from the ejecta launched by the merger (a mechanism initially theorised by \citealt{1976ApJ...210..549L} for black hole -- neutron star mergers, extended by \citealt{1998ApJ...507L..59L, 1999A&A...341..499R} and \citealt{PhysRevD.87.024001} to binary neutron star (NS) mergers and with spectra and light curves calculated by \citealt{10.1111/j.1365-2966.2010.16864.x, 2013ApJ...774...25K}; see also \citealt{MetzgerLR} for a living review on kilonovae). The KN is generated by the radioactive decay of the neutron-rich matter through r-process decay, leading to an approximately power-law decrease in the bolometric luminosity and effective temperature of the KN during the deceleration phase of the GRB afterglow \citep{10.1111/j.1365-2966.2010.16864.x, Troja211211A}. The peak luminosity of the KN can vary depending on the neutron heating \citep{MetzgerKNprecursor} and lanthanide richness of the ejecta \citep{Barnes_2013, 10.1093/mnras/stu802}.  A lanthanide-rich, or "red", KN will delay (and lower) the peak emission to a timescale of $\sim 1$ week due to the delay in the ejecta becoming optically thin, while lanthanide-poor, or "blue", KN will peak around a day or less. If neutron heating is included, the timescale can go down to hours or even sub-hour, depending on the model parameters used (though there is ongoing debate on the modelling of neutron heating; see \citep{Sarin_2024}). The first (tentative) kilonova detected was found with the short GRB 130603B \citep{2013Natur.500..547T}, and the connection between kilonovae and GRBs was firmly established with the detection of AT 2017gfo alongside GRB 170817A \citep{Abbott_2017_2, 2018MNRAS.481.3423W}.

Since the kilonova is (quasi-)isotropic, any emission that reaches the observer must also pass through the GRB blast-wave, such that the KN acts as a source of seed photons around the optical regime \citep{KilonovaSeedPhotons}. If the emission from the KN exceeds the optical emission from the synchrotron emission of the forward shock, then it is possible that the IC emission will be impacted. \citet{2022Natur.612..236M} considered this scenario for GRB 211211A, but disfavoured it due to the forward shock synchrotron emission dominating over the kilonova component, leading to an SSC spectral peak that was too energetic. Instead, they argued that the up-scattering occurred in a late low power jet nearer to the kilonova itself. However, in sGRBs, the synchrotron emission is generally weaker than long GRBs, allowing for more scenarios where the KN seed photons can dominate at earlier times and modify the IC emission.

In both sGRBs and long GRBs, the energy of the jet is expected to have some angular dependency, leading to structured jets (see for example \citealt{Mészáros_Struct, Rossi2002, Granot_2003, Zhang_2004}). While this was first considered soon after the discovery of GRB afterglows (\citealt{costa1997discovery, van1997transient}), it was not until GRB 170817A \citep{PhysRevLett.119.161101, Abbott_2017} that an afterglow was observed that would not fit the usual top-hat\footnote{Flat energy distribution with angle.} model and instead was better fitted by an off-axis structured jet (as outlined in papers such as \citealt{2017Natur.551...71T, Troja2018, 170817AX-ray, Alexander_2018, Wu_2018, Troja2019}). Since it is also the first confirmed GRB associated with a kilonova \citep{Abbott_2017_2, 2018MNRAS.481.3423W}, GRB 170817A is a particularly interesting candidate for studying the impact of KN seed photons on its VHE emission. No VHE emission was detected in this case \citep{Abdalla_2017}, likely due to a variety of effects including Klein-Nishina effects, the reduction of flux due to off-axis viewing and, when considering its TeV emission, extragalactic background light (EBL) attenuation \citep{Clement, MyPaper}. Nonetheless, it may be instructive to consider such a case to understand the impact of KN seed photons, especially as further off-axis sGRBs are expected to be observed in the future, along with their jet structure. This varying jet structure has angularly dependent dynamical properties that can be affected differently by the KN emission. Additionally, there is an expected increase to VHE GRBs (short or long) due to the increased sensitivity of the upcoming CTA observatory \citep{2019scta.book.....C}.

In this paper, we consider the impact of the KN on the afterglow of sGRBs. The afterglow is modelled dynamically using a shell model, with the electron-photon gas evolved using a kinetic solver in a similar fashion to \citet{MyPaper}. We inject a thermal photon population into this shell that models the seed photons arriving from the KN. From this, we can infer the impact of the KN seed photons on the VHE emission. We find that they are a promising seed photon source which can have a substantial impact on VHE emission, though due to the weakness of sGRBs and effects such as EBL attenuation, their impact can be difficult to observe.

The remainder of this paper is structured as follows: in section~\ref{sec:Method}, we outline the steps taken to model the GRB afterglow and KN seed photon injection, through the combined use of a kinetic code and shell model, and with flux calculated with respect to the equal arrival time surface. In section~\ref{sec:Params}, we outline the impact of the KN seed photons on the VHE emission as we vary the microphysical parameters and jet energy. In section~\ref{sec:Jet} we fix the afterglow parameters and vary the jet structure and observer angle to see how the VHE emission with and without KN seed photons would be observed. In section~\ref{sec:RealGRBs}, we model the VHE afterglow of GRB 170817A with and without KN seed photons to see what impact the KN may have had on the TeV emission. In section~\ref{sec:Discuss}, we discuss further how the KN seed photons impact the IC emission, and we outline an analytical approach to assess whether KN seed photons should be accounted for when modelling VHE emission. In section~\ref{sec:Conclude} we outline our conclusions.

Throughout this paper, we assume a standard $\Lambda$CDM cosmology with $H_{0} = 67.4$, $\Omega_{m} = 0.315$ and $\Omega_{\Lambda} = 0.685$.

\section{Method}
\label{sec:Method}
In this section, we start by briefly outlining the kinetic code, shell model and flux calculations used in this work, which is based off the approach used in \citet{MyPaper}, though with the baryonic--loading in this work fixed for all angles. The approach used for KN seed photon injection is also presented.

\subsection{Kinetic modelling of structured GRBs}
\subsubsection{The kinetic code}
For calculating the self-consistent evolution of the electron and photon gas, we use the kinetic code \textsc{katu} \citep{Katu1, Katu2}, with modifications made to it for the inclusion of adiabatic expansion and general IC cooling of electrons \citep{MyPaper}. \textsc{katu} solves a series of kinetic equations, given in the general form by
\begin{equation}
    \frac{\partial n_X}{\partial t^{\prime}} = Q_{e} + Q_{i} + \mathcal{L} - \frac{n_X}{\tau_{\rm esc}} - \frac{\partial}{\partial\gamma_{e}}\left(\skew{-6}\dot{\gamma_{e}}n_X\right), \label{eq:Katu_kinetic}
\end{equation}
where $n_X$ is the particle population of a particle species $X$, $t^{\prime}$ is the time in the fluid frame, $Q_{e}$ is the external injection, $Q_{i}$ is the internal injection, $\mathcal{L}$ is the internal losses, $\tau_{\rm esc}$ is the escape timescale\footnote{This is only applied to the photon population} and the last term represents losses due to synchrotron and IC cooling. We add to this adiabatic cooling\footnote{This only impacts the electron population -- the photon escape timescale is always smaller than its adiabatic timescale, ensuring adiabatic expansion can be neglected for the photon population.} such that
\begin{equation}
    \frac{\partial n_{e}}{\partial t^{\prime}}\bigg|_{\rm adb} = -\frac{n_{e}}{\tau_{\rm ad}} - \frac{\left(1-\hat{\gamma}\right)}{\tau_{\rm ad}}\frac{\partial}{\partial \gamma_{e}}(n_{e}\gamma_{e}),
    \label{eq:Adb_total}
\end{equation}
where $\hat{\gamma}$ is the adiabatic exponent and $\tau_{\rm ad}$ is the adiabatic timescale in the fluid frame. $\tau_{\rm ad}$ is defined as
\begin{equation}
    \frac{1}{\tau_{\rm ad}} = 3\gamma\frac{\skew{-11}\dot{R_{\rm sh}}}{R_{\rm sh}} -\dot{\gamma},
\end{equation}
where $\gamma$ is the fluid Lorentz factor and $R_{\rm sh}$ is the radius of the forward shock, which evolves as
\begin{equation}
    \frac{dR_{\rm sh}}{dt} = c\beta_{\rm sh},
\end{equation}
for a shock with instantaneous velocity $\beta_{\rm sh}$. We assume the electron population is relativistic throughout, and therefore set $\hat{\gamma} = 4/3$.

IC cooling is modelled with the inclusion of variation to the cross-section due to the photon/electron energy, such that \citep{JonesIC, BlumGould}
\begin{equation}
\label{eq:IC_gen_cool}
    \frac{d\gamma_{\rm ic}}{dt} = - \int_{\epsilon_{\rm min}}^{\epsilon_{\rm max}}n_{\gamma}(\epsilon)\int_{\epsilon}^{\frac{4\epsilon\gamma_{e}^{2}}{1 + 4\epsilon\gamma_{e}}}\epsilon_{s}\sigma_{\rm KN}(\gamma_{e}, \epsilon, \epsilon_{s})d\epsilon_{s}d\epsilon,
\end{equation}
where $\epsilon$ and $\epsilon_{s}$ are the pre and post-scattered photon energies normalised to the rest mass energy of an electron, $\gamma_{e}$ is the electron's particular Lorentz factor in the fluid frame and $n_{\gamma}(\epsilon)$ is the pre-scattered photon population. $\sigma_{\rm KN}(\gamma, \epsilon, \epsilon_{s})$ represents the full Klein-Nishina cross-section, and is given by
\begin{multline}
    \sigma_{\rm KN}(\gamma_{e}, \epsilon, \epsilon_{s}) = \frac{3c\sigma_{T}}{4\gamma_{e}^{2}\epsilon}\left[\vphantom{\frac{\epsilon_{s}^{2}}{\gamma(\gamma - \epsilon_{s})}} 2q_{u}\ln q_{u} + (1 + 2q_{u})(1 - q_{u})\right. \\ 
    \left. + \frac{\epsilon_{s}^{2}}{\gamma_{e}(\gamma_{e} - \epsilon_{s})}\frac{1 - q_{u}}{2}\right]
\end{multline}
where $q_{u} = \frac{\epsilon_{s}}{4\epsilon\gamma_{e}(\gamma_{e} - \epsilon_{s})}$, for $q_{u} \leq 1 \leq \frac{\epsilon_{s}}{\epsilon},~\epsilon_{s} < \gamma_{e}$.

The electron population is then solved with the inclusion of a non-thermal population $Q_{e} \propto \gamma_{e}^{-p}$ injected from the forward shock due to Fermi-I acceleration (see for example, \citealt{2001MNRAS.328..393A, Sironi_2013}), and cooled by adiabatic, IC cooling and synchrotron cooling.

\subsubsection{The shell model}
To determine the dynamical evolution of the GRB jet, we use a shell model, which has seen extensive use in GRB modelling (see for example, \citealt{10.1093/mnras/stt872, 2013arXiv1309.3869V, HJvERev, afterglowpy}). This is based on the fireball model (see e.g. \citealt{ReesMeszaros1992, MeszarosRees1993, 1995ApJ...455L.143S, PanaitescuKumar2000, Meszaros2002}), with the acceleration phase neglected as it occurs before the initial observed afterglow emission.

The GRB is treated as a thin spherical shell expanding outwards with a Lorentz factor $\gamma \gg 1$ (or velocity $\beta \sim 1$) and isotropic-equivalent energy $E_{\rm iso}$. This is a combination of the initial kinetic energy in the ejecta mass $E_{\rm ej,~iso}$ and the energy acquired by the circumburst matter swept-up by the shell $E_{\rm sw,~iso}$, such that
\begin{equation}
    \label{eq:E_iso}
    E_{\rm iso} = E_{\rm ej,~iso} + E_{\rm sw,~iso}
\end{equation}
where
\begin{equation}
    \label{eq:E_ej}
    E_{\rm ej,~iso} = (\gamma - 1) M_{\rm ej}c^{2},
\end{equation}
with $M_{\rm ej}$ being the ejecta mass, and
\begin{equation}
    \label{eq:E_sw}
    E_{\rm sw,~iso} = \frac{\beta^{2}}{3}(4\gamma^{2} - 1)M_{\rm sw}c^{2},
\end{equation}
where $M_{\rm sw}$ is the swept-up mass. This leads to an initial coasting phase when $E_{\rm ej,~iso}$ is dominant, followed by a deceleration phase when $E_{\rm sw,~iso} \gtrsim E_{\rm ej,~iso}$.

The width of the shell is given by
\begin{equation}
    \label{eq:Delta_R}
    \Delta R = \frac{R_{\rm sh}}{4(3 - k)\gamma^{2}},
\end{equation}
which leads to a photon escape timescale
\begin{equation}
    \tau_{\rm esc} = \frac{\pi}{4}\frac{\gamma \Delta R}{c},
    \label{eq:tau_esc}
\end{equation}
where we have assumed that the region from which photons escape is locally cylindrical. $\tau_{\rm esc}$ also sets the upper bounds of the timestep $\Delta t_{\rm fluid} = \Delta t_{\rm lab} / \gamma$ for the kinetic solver due to the stiffness of the equations.

Since the front of the shell will produce a strong relativistic shock, and assuming an equation of state such as that outlined in \citet{2013arXiv1309.3869V}, we can obtain the shock-jump conditions that govern the fluid mass density $\rho$ and internal energy density $e$:
\begin{align}
    \rho &= 4\gamma\rho_{\rm ext}c^{2}, \\
    e &= 4\gamma\left(\gamma - 1\right)\rho_{\rm ext}c^{2},
\end{align}
where $\rho_{\rm ext}$ is the external mass density. The shock Lorentz factor $\Gamma_{\rm sh}$ can be related to $\gamma$ by

\begin{equation}
    \label{eq:Gamma_sh}
    \Gamma_{\rm sh}^2 = \frac{(4\gamma^{2} - 1)^{2}}{8\gamma^{2} + 1} \sim 2\gamma^{2},
\end{equation}
for $\gamma >> 1$.

The B-field strength $B$ is driven by the internal density, such that
\begin{equation}
    \label{eq:B-field}
    \frac{B^{2}}{8\pi} = \epsilon_{B}e,
\end{equation}
where $\epsilon_{B}$ is the fraction of internal energy given over to the B-field.

The injected population of electrons from the shock front are bounded in energy by $\gamma_{\rm min}$ and $\gamma_{\rm max}$. $\gamma_{\rm min}$ is set by normalising the injected population, and is given by

\begin{equation}
\label{eq:gamma_min}
    \gamma_{\rm min} = \frac{p - 2}{p - 1}\frac{\epsilon_{e}e}{nm_{e}c^{2}}, ~(p > 2)
\end{equation}
where $\epsilon_{e}$ is the fraction of the internal energy density given to the electrons and $n$ is the total number density of electrons in the post-shock fluid.

$\gamma_{\rm max}$ is set by equating the rate of synchrotron cooling
\citep{Granot_2002}
\begin{equation}
    \skew{-5}\dot{\gamma_{e}} = - \frac{\sigma_{T}B^{2}}{6\pi m_{e}c}\gamma_{e}^{2},
\end{equation}
with the acceleration timescale as estimated by the relativistic gyro-frequency $\omega_{B} = q_{e}B/\gamma_{e}m_{e}c$, such that \citep{Dai_2001, 2009A&A...498..677B, MyPaper}
\begin{equation}
\label{eq:gamma_max}
    \gamma_{\rm max} = \eta\sqrt{\frac{3q_{e}}{\sigma_{T}}}B^{-\frac{1}{2}} \approx 4.65 \times 10^7 \eta \left( \frac{B}{1 \enskip \rm{ Gauss}}\right)^{-\frac{1}{2}},
\end{equation}
where $q_{e}$ is the elementary charge and $\eta$ is a scale factor of order unity that can be adjusted to account for the uncertainty in this approximate assumption. For this work we set $\eta$ to $1$.

The bulk Lorentz factor $\gamma$ and distance to the origin $R_{\rm sh}$ are jointly solved with a 4th order Runge-Kutta routine, using the time differentiated form of eq.~\ref{eq:E_iso} and the fact that $\skew{-11}\dot{R_{\rm sh}} = c\beta_{sh}$.

\subsubsection{The structured jet}
To model the structure of the Gaussian and power-law jets, we subdivide the jet into a series of independent top-hat jets (zones) with equal solid angles. The number of zones must be set sufficiently high that the structure is reproduced by this approximation. We found in \citet{MyPaper} that $18$ zones was a good compromise between accuracy and computational cost and so used $18$ for this work. However, if the initial Lorentz factor of a particular zone falls below $2$, we discount the zone as its output is negligible compared to the rest of the jet.

Structured jets have been in literature for a long time (see for example, \citealt{10.1046/j.1365-8711.2002.05363.x, Mizuta_2009, Duffell_2013, 10.1093/mnras/stx2345, Lazzati_2017}) as a means to capture simulation results. However, with the advent of GRB 170817A, there has been an increased interest in using structured jets to describe both long and short GRBs (such as in \citealt{Margutti_2018, Geng_2019, DuqueJet, 2020A&A...636A.105S, 2020MNRAS.497.1217T, 10.1093/mnras/stab032, 2022Univ....8..612L, Ren_2024}). For this work, we make use of structured jets based on the parameterised models found in \citet{afterglowpy}, which share normalised jet energy $E_{0}$ and normalised core width $\theta_{c}$. For an isotropic equivalent energy $E_{\rm iso} \equiv E(\theta) = 4\pi dE/d\Omega$, the energy distributions of the jets are given by
\begin{align}
\label{eq:StructTop}
    E (\theta) &= E_{0} ~~~~~~~~~~~~~~~~~~~~~~~~~~~~~~~\text{Top-hat,} \\
\label{eq:StructGauss}
    E (\theta) &= E_{0} \exp\left(-\frac{\theta^{2}}{2\theta_{c}^{2}}\right) ~~~~~~~~~~\text{Gaussian,} \\
\label{eq:StructPower}
    E (\theta) &= E_{0} \left(1 + \frac{\theta^{2}}{b\theta_{c}^{2}}\right)^{-b/2} ~~~~~\text{Power-law,}
\end{align}
where $E_{0}$ is the normalised energy ($E_{\rm iso}$ at $\theta = 0$) and $b$ is the power-law index.

Because we have a coasting phase, we must also account for the ejecta mass and initial Lorentz factor. In this work, we model this by using the same distribution as $E_{\rm iso}$ to determine the initial Lorentz factor $\gamma$ (e.g. for the Gaussian jet, $\gamma = \gamma_{0}\exp\left(-\theta^{2}/2\theta_{c}^{2}\right)$, with $\gamma_{0}$ being the normalised Lorentz factor at $\theta = 0$; the jet is truncated for $\gamma < 2$ to avoid unphysical results.). The Baryon-loading of the ejecta along a given angle is set by the combination of $\gamma$ and $E$, under the assumption that the pre-deceleration/coasting outflow is relativistically cold, as expressed in Equation \ref{eq:E_ej}.

\subsubsection{Calculating the observed flux}
The simulated photon population $n_{\rm ph}(\epsilon)$ is converted to a power radiated per unit volume per unit frequency (in the fluid frame)
\begin{equation}
    P_{\nu} = h\frac{n_{\rm ph}\epsilon_{\rm ph}}{\tau_{\rm esc}},
\end{equation}
which is then used to calculate the flux by integrating over the solid angle of a particular isotropic zone of the jet,
\begin{equation}
    \label{eq:FluxGen}
    F_{\nu}(t_{\rm obs}, \nu_{\rm obs}) \approx \frac{1 + z}{4\pi d_{\rm L}^{2}}P_{\nu}\int_{\rm zone} d\Omega dr~r^{2}\delta^{2},
\end{equation}
where $z$ is the cosmological redshift, $d_{\rm L}$ is the luminosity distance, $\Omega$ is the solid angle, $r$ is the distance from origin for the emitting zone and $\delta$ is the Doppler boosting factor (see e.g. \citealt{vanEerten2010, afterglowpy}). While this equation assumes an optically thin medium, self-absorption is accounted for by \textsc{katu} during the simulation of the photon population \citep{Katu2}.

Due to the homogeneity and isotropy of the zone, $P_{\nu}$ is not dependent on radius or solid angle and may be taken as fixed for a given lab time at any point in that zone. The entire contribution of the blast-wave over the equal arrival time surface is then summed by dividing the zones into point source segments with a given $\left(\theta, \phi\right)$, added together to get the total flux for a given zone. For this work, we found a segment number (for the entire jet) of $(1500, 1)$ for on-axis ($\theta_{obs} = 0$) and $(800,400)$ for off-axis\footnote{We halve the number of $\phi$ points by taking advantage of the axi-symmetric nature of the jet.} runs was a good balance between resolved flux and computational resources.

In the observer frame, the width of the shell appears extended due to the shell almost keeping up with prior emission, which gives us a width 
\begin{equation}
    \label{eq:delta_R}
    dr \equiv \Delta R_{\rm obs} = \frac{\Delta R}{1 -\beta_{\rm sh}\mu},
\end{equation}
where $\mu$ is the cosine of the angle between an emitting region and the observer, which for a given $(\theta, \phi)$ is
\begin{equation}
    \label{eq:mu}
    \mu = \sin\theta\cos\phi\sin\theta_{\rm obs} + \cos\theta\cos\theta_{\rm obs}.
\end{equation}
Each segment of the jet will be seen by the observer at a time
\begin{equation}
\label{eq:t_obs}
    t_{\rm obs} = \left(1 + z\right)\left(t_{\rm lab} - \frac{R_{\rm sh}\mu}{c}\right).
\end{equation}

The radial extent was captured by subdividing the lab width into $K = 100$ regions along $\mathbf{\hat{r}}$, with the $k$th point source given by
\begin{equation}
\label{eq:radial_integ}
    r_{k} = R_{\rm sh} + \frac{\Delta R}{2K}(1 - 2(k + 1)), ~~ 0 \leq k \leq K.
\end{equation}

EBL attenuation was also included \citep{1992ApJ...390L..49S, doi:10.1126/science.1227160}, which can substantially affect the VHE emission for $z \gtrsim 0.08$ in GRBs \citep{Stecker_2006, 10.1093/mnras/stad1388}. This can be modelled by setting the flux as
\begin{equation}
    F_{\nu,\rm~obs} = F_{\nu,\rm~init} \times e^{-\tau(\nu, z)},
\end{equation}
where $F_{\nu,\rm~init}$ is the flux as given by eq.~\ref{eq:FluxGen} and $\tau(\nu, z)$ is the optical depth of the EBL attenuation \citep{doi:10.1146/annurev.astro.39.1.249}. We have applied the model provided by \citet{10.1111/j.1365-2966.2010.17631.x} to find $\tau(\nu, z)$.

\subsection{Modelling of a Kilonova source}
We apply a simplified model of the KN emission by assuming that it is a blackbody emitter at origin with a bolometric luminosity $L (t)$ and effective temperature $T(t)$. Here, $t$ is the emission time of the light leaving the emitter, given by
\begin{equation}
    t = t_{\rm lab} - R_{\textrm{back}}/c,
\end{equation}
where $t_{\rm lab}$ is the arrival time and $R_{\textrm{back}}$ is the radius of the back of the shell for a given zone, $R_{\rm back} = R_{\rm sh} - \Delta R$. If redshift is ignored, $t$ and $t_{\rm obs}$ can be conflated.

While in reality the KN will have a range of temperatures across it due to the various ejecta shells having different cooling rates, velocities and opacities,  using a single effective temperature provides us with a good approximation of the overall KN \citep{Sneppen_2023}. Furthermore, though the KN is a spherically expanding shell, its relatively small radius compared to the afterglow shell allows us to treat the KN approximately as a point source (in the example of 211211A, $R_{KN} \sim 10^{14}-10^{15}$ cm \citep{Troja211211A} while $R_{\rm GRB} \sim 10^{16}-10^{17}$ cm).

The bolometric luminosity and effective temperature broadly follow a power-law once KN emission begins to decline post-peak (see for example \citealt{2017Natur.551...75S, MetzgerLR, Troja211211A}). We can thus set the luminosity and temperature as
\begin{align}
    L &= L_{0}\left(\frac{t}{t_{0}}\right)^{-\alpha_{L}}, \\
    T &= T_{0}\left(\frac{t}{t_{0}}\right)^{-\alpha_{T}}, 
\end{align}
where $L_{0}$ and $T_{0}$ are normalisation factors for reference time $t_{0}$ and $\alpha_{L}$ and $\alpha_{T}$ are their respective power-law indices. The values for $L_{0}, ~T_{0}, ~\alpha_{L}$ and $\alpha_{T}$ can be obtained by fitting a power-law function against the measurements of the bolometric luminosity and temperature from KN associated with GRBs such as GRB 211211A \citep{Troja211211A} and GRB 170817A \citep{2017Natur.551...75S}. For this work, we set $t_{0} = 30$ mins and calibrate $L_{0}$ and $T_{0}$ accordingly.

From this, we can obtain an injected photon distribution at time $t_{\rm lab}$, which is given by
\begin{align}
    \frac{dn(\epsilon)}{dt^{\prime}} =& 
    \left(\frac{m_{e}c^{2}}{h}\right)^{3}
    \frac{2{\epsilon}^{2}}{V^{\prime}c^{2}}
    \frac{\gamma^{4}\left(1 + \beta\right)^{4}}{\exp(\frac{\gamma(1 + \beta)m_{e}c^{2}\epsilon}{k_{B}T\left(t\right)}) - 1} \notag \\
    &\times \frac{\pi L\left(t\right)}{\sigma T\left(t\right)^{4}}(1 - \beta_{\rm back}), \label{eq:KNInj}
\end{align}
where $V^{\prime}$ is the volume of the shell in the fluid frame, $h$ is the Planck constant, $k_{B}$ is the Boltzmann constant and $\left(1 - \beta_{\rm back}\right)$ is a factor that accounts for the loss of flux due to the shell receding from the kilonova source. $\beta_{\rm back}$ is the velocity of the rear of the shell, which varies slightly from $\beta$ as the shell width also increases in time. By accounting for the time evolution of $R_{\rm sh}$ and $\Delta R$, it follows that
\begin{equation}
    \beta_{\rm back} = \frac{1}{c}\left( \skew{3}\dot{R}\left(1 - \frac{1}{12\gamma^{2}}\right) + \frac{\skew{-16}\dot{\Gamma_{sh}}R}{6\gamma^{3}}\right).
\end{equation}

Because the power-law will become unphysically high at early times, the earliest emission time allowed for the KN to impact the shell is set to $30$ minutes in the observer frame, before which we assume that there is some initial rising power-law luminosity (see \citealt{MetzgerKNprecursor, MetzgerLR}). This part of the kilonova is not modelled for this work.
The derivation of eq.~\ref{eq:KNInj} is given in appendix~\ref{App:KN}.

\section{The role of afterglow parameters}
\label{sec:Params}

To start, we consider a standard top-hat, on-axis GRB, where we have tuned the parameters in our model to represent a sGRB. We set $E_{0} \sim 10^{51}$ ergs and $n_{\rm ext} \sim 10^{-3}~\rm cm^{-3}$ as average values for sGRBs, and set $\epsilon_{e}$\footnote{Which is normally of order $0.1$, see \citet{10.1093/mnras/stac246}.} and $\epsilon_{B}$\footnote{Which can vary from $10^{-5}-10^{-1}$, see \citet{10.1093/mnras/stad3272}.} to $10^{-1}$ and $10^{-4}$ respectively as reasonable quantities \citep{2015ApJ...815..102F}. $z$ is set to $0.64$ as the median cosmological redshift for sGRBs \citep{Nugent_2022}. For our assumed cosmological parameters, this corresponds to a luminosity distance $d_L$ of $1.21 \times 10^{28}$ cm. $\gamma_{0}$ \citep{Liang_2010}, $\theta_{c/w}$, $b$ and $\chi_{N}$ \citep{afterglowpy} and $p$ \citep{2001MNRAS.328..393A, afterglowpy} were chosen as expected parameters for the GRBs in question. The bolometric luminosity and temperature of the KN used in these runs are based on the measured values for GRB 211211A as given in \citet{Troja211211A}. It has been recently suggested by \citet{2025ApJ...984...37W} that the secondary emission associated with GRB 211211A is actually thermal radiation from the dust surrounding the GRB being heated by UV and soft-X-ray light produced by the GRB acting on the circumstellar medium (CSM), as opposed to a KN \citep{Troja211211A, 2022Natur.612..223R, 2022Natur.612..232Y}. In either case, the similarity of the emission to confirmed KN such as GRB 170817A (as noted by \citet{Troja211211A}) means that the following results and discussion are agnostic to this issue. The full set of baseline sGRB values are given in Table~\ref{tab:24_run_params}.

\begin{figure*}
    \centering
    \includegraphics[width=0.85\linewidth]{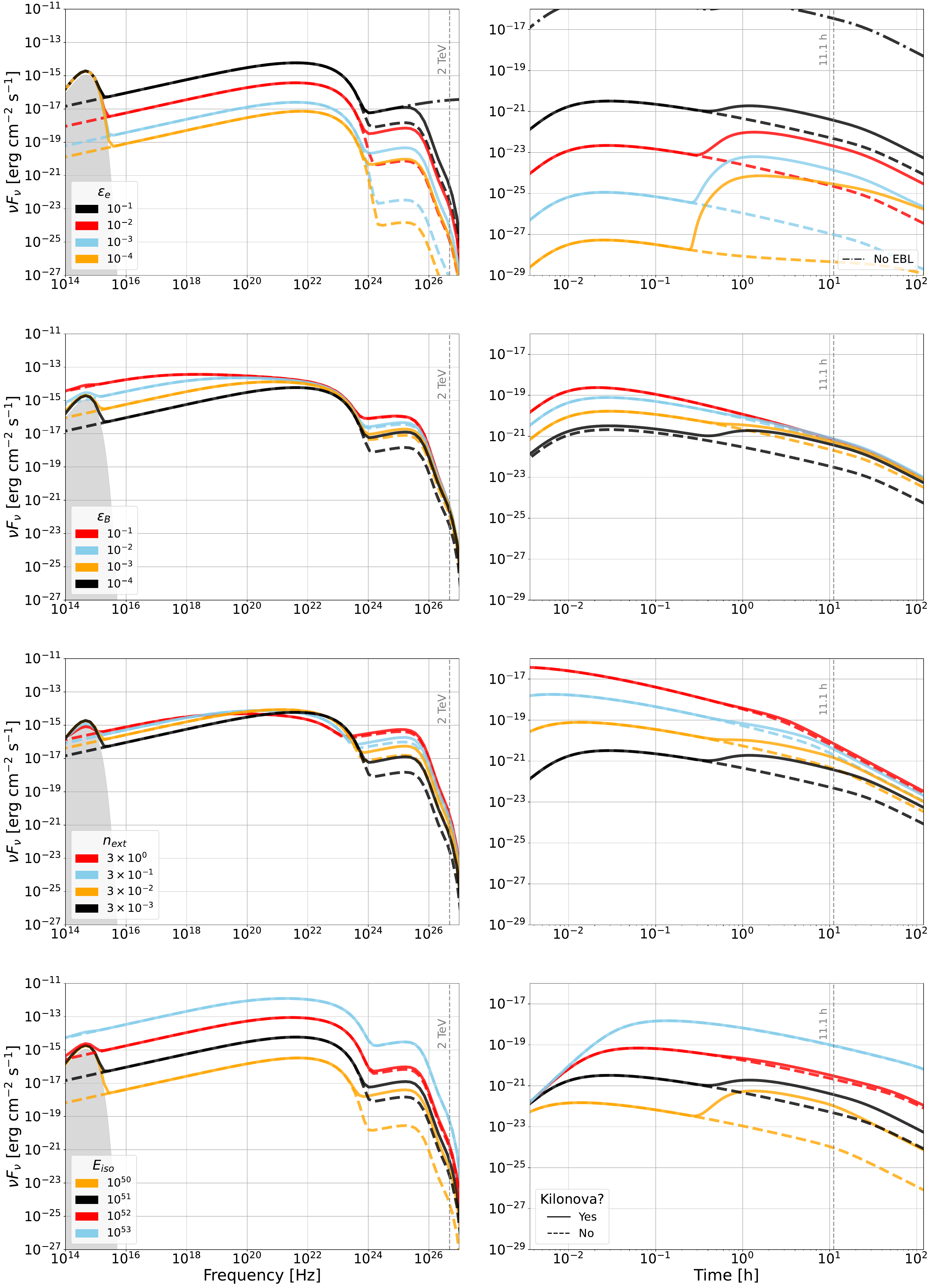}
    \caption{Spectra at $11.1$ hours and $4.84 \times 10^{26}$ Hz (2 TeV) light curves for the 4 top-hat scenarios where $\epsilon_{\rm e}$ (row $1$), $\epsilon_{\rm B}$ (row $2$), $n_{\rm ext}$ (row $3$) and $E_{\rm iso}$ (row $4$) are varied. Runs with KN seed photon injection (solid lines) and with no such injection (dashed lines) are overlaid for comparison. For the $\epsilon_{e} = 10^{-1}$ case, we also provide a KN scenario where EBL attenuation is neglected (dot-dashed lines). The independent KN emission at the observed time used in the spectra is given by the shaded area. The $11.1$ h and $2$ TeV positions are marked using a grey dashed line on the light curves and spectra, respectively. All results are given in units of $\nu F_{\nu}$ ($\mathrm{erg~cm^{-2}~s^{-1}}$), with frequency in Hz and observer time in days. Unless stated otherwise in the corresponding legend, the parameters used are from table~\ref{tab:24_run_params}.}
    \label{fig:Series_plot}
\end{figure*}

We then vary $\epsilon_{e},~\epsilon_{B},~n_{\rm ext}$ and $E_{0}$ to observe the impact on the observed spectra and VHE emission, as compared to the no-KN scenario. The results from this are shown in Fig.~\ref{fig:Series_plot}. It should be noted that, if the spectral regime of the afterglow is known, it can be analytically inferred if the KN will have an impact on the VHE emission. We leave this point to our discussion in section~\ref{sec:Constraint}.

Overall, the impact of the KN is driven by whether its flux is able to dominate the near-IR, optical or UV regime (depending on observer time) of the forward shock, with the largest impact seen in the cases where the synchrotron emission of the forward shock is significantly suppressed. Of particular note is that, due to EBL attenuation (compare the case of $\epsilon_{e}$ with and without EBL attenuation in the top row of Fig.~\ref{fig:Series_plot}; see also sec.~\ref{sec:Impact}) the peak flux of the IC emission is only on the order of $\sim 100 ~\rm GeV$. Combined with Klein-Nishina suppression (especially at later times where Doppler boosting is weaker; see also \citealt{NakarKN, 10.1093/mnras/stab911, McCarthy_2024}), this leads to a severe reduction in IC emission, which impacts both the up-scattered KN and SSC emission. In each parameter test, the scenarios with the brightest synchrotron emission sees the strongest VHE emission, with less impact from the KN, while in the cases of weaker synchrotron emission, the KN seed photons help to boost the VHE emission and help the emission overcome the effects of EBL attenuation and Klein-Nishina suppression.

This is particularly noticeable in the case of $\epsilon_{e}$ (top row of Fig.~\ref{fig:Series_plot}). By decreasing $\epsilon_{e}$, we reduce the synchrotron emission, since for $\nu_{m} < \nu < \nu_{c}$, $F_{\nu} \propto \epsilon_{e}^{p-1}$ \citep{Granot_2002}. However, the reduction in $\epsilon_{e}$ also decreases of number of photons up-scattered to very-high energies \citep{NakarKN}. By injecting seed photons from the KN, this reduction in up-scattering is mitigated by the increased number of seed photons available for scattering around $10^{15}$ Hz. This can lead to, as seen for $\epsilon_{e} = 10^{-4}$, an increase of around $4$ orders of magnitude in the VHE emission, causing the TeV emission to become equivalent to, and at late times along with $\epsilon_{e}\sim10^{-3}$, exceed the emission with no-KN seed photons from the $\epsilon_{e} = 10^{-2}$ case. In all $\epsilon_{e}$ cases except $10^{-1}$, the KN seed photons boost the peak TeV higher and later than what would otherwise be expected if only considering the forward shock. In the case of $10^{-1}$, a second, slightly lower peak in the TeV emission is produced, leading to a boost of $\sim 1$ order of magnitude in the TeV emission over the forward shock emission.

\begin{table}
\caption{Baseline afterglow and KN parameters used for the sGRB runs in figures \ref{fig:Series_plot} and \ref{fig:PR2_plot}}
\label{tab:24_run_params}
\centering
\begin{tabular}{lccc}
\hline
Parameter & Value & Description \\
\hline
$E_{0}$ & $10^{51}$ erg &  Jet isotropic equivalent energy \\
$\theta_{c}$ & $4.58$\textdegree ~ ($0.08$ rad) & Jet core angle \\
$\theta_{w}$ & $13.75$\textdegree ~ ($0.24$ rad) & Wing truncation angle \\
 &  & (power-law/Gaussian only) \\
$n_{\rm ext}$ & $0.003$ cm$^{-3}$ & External number density\\
$\gamma_{0}$ & $200$ & Initial coasting Lorentz factor\\
$R_{\rm fire}$ & $10^{8}$ cm & Fireball radius\\
z & $0.64$ & Cosmological redshift \\
$d_{L}$ & $1.21 \times 10^{28}~\rm cm$ & Luminosity distance \\
p & $2.2$ & Spectral slope\\
b & $3$ & Power-law structure slope \\
 & & (power-law only) \\
$\epsilon_{e}$ & $10^{-1}$ & Electron energy fraction\\
$\epsilon_{B}$ & $10^{-4}$ & B-field energy fraction\\
$L_{0}$ & $5.57 \times 10^{43} \rm ergs^{-1}$ & Normalised luminosity \\
$T_{0}$ & $3.52 \times 10^{4} \rm ~K$ & Normalised temperature \\
$t_{0}$ & 1800 s & Reference time for $L_{0},T_{0}$ \\
$\alpha_L$ & $1.148$ & Luminosity power-law index \\
$\alpha_T$ & $0.497$ & Temperature power-law index \\
$\chi_{N}$ & $1$ & Fraction of accelerated electrons\\
$\eta$ & $1$ & $\gamma_{\rm max}$ scale factor\\
\hline
\end{tabular}
\end{table}

When increasing $\epsilon_{B}$ ($2$nd row of Fig.~\ref{fig:Series_plot}), we boost the synchrotron emission, causing the KN seed photons to become subdominant to the synchrotron photons. The KN injected $\epsilon_{B} = 10^{-4}$ case is comparable to the other scenarios beyond $1$ day as the TeV light curve is further constrained by the higher $\epsilon_{B}$ value. We can also see that, due to the combined effects of low $n_{\rm ext}$, $E_{\rm iso}$ and $\epsilon_{B}$, that the ratio of IC to synchrotron emission is reduced at lower $\epsilon_{B}$ compared to the synchrotron. This is primarily due to two effects; the EBL attenuation of the TeV emission as mentioned above, and in this particular case the high cooling break $\nu_{c}$, which is boosted by the weak magnetic field generated at low $\epsilon_{B}$ and $n_{\rm ext}$. This is further exacerbated at early times when the cooling break can exceed the synchrotron cut-off frequency. Klein-Nishina suppression can also be strengthened by lower $\epsilon_{B}$ values, further reducing IC scattering \citep{NakarKN}.

Increasing $n_{\rm ext}$ (see third row of Fig.~\ref{fig:Series_plot}) lowers $\nu_{c}$, leading to higher VHE emission, but quickly suppresses the impact of the KN seed photons such that the TeV light curves for the KN and no-KN cases at $n_{\rm ext} \gtrsim 1~\rm cm^{-3}$ show minimal deviation from each other. At this stage, the peak TeV emission is then fully controlled by the synchrotron photons. Nonetheless, at low $n_{\rm ext}$, the KN seed photons can boost the TeV emission so that beyond $1$ day the $4$ cases are within an order of magnitude to each other.

Finally, varying $E_{\rm iso}$ (bottom row, Fig.~\ref{fig:Series_plot}) shows a similar effect to varying $\epsilon_{e}$, with the KN $10^{50}$ case having its peak TeV emission boosted and occurring later in time, such that it slightly exceeds the no-KN $10^{51}$ case from 0.8 to 30 hours before converging due to the earlier jet break in the $10^{50}$ case. The rise in the light curves due to the KN seed photons reaching the shell is flatter and smoother than the ones obtained when varying $\epsilon_{e}$, due to the reduced impact of the KN in the low $E_{\rm iso}$ cases compared to the low $\epsilon_{e}$ cases. For $E_{\rm iso} \gtrsim 10^{52}$, the impact of the KN with the luminosity and temperature used in these runs is negligible.

\begin{figure*}
    \centering
    \includegraphics[width=0.98\linewidth, trim = {0 0 0 0}]{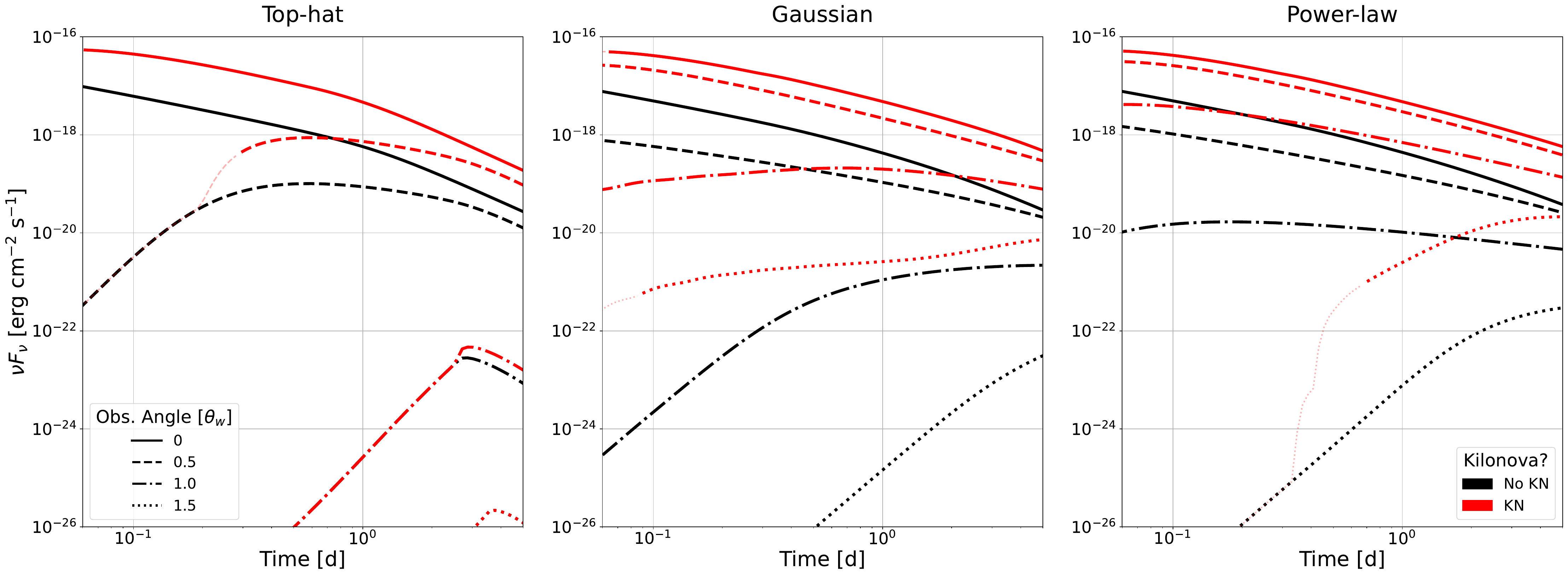}
    \caption{$100$ GeV light curves for the 12 structured and observer angle scenarios. The three jet structures are top-hat (left), Gaussian (middle) and power-law (right), with $4$ observer angles ($0,~0.5,~1$ and $1.5$ times the half-opening angle of the jet). Results with and without seed photons from the KN are shown together. Where light curves are faint, the result is impacted by the choice of start time for the KN emission. All results are given in units of $\nu F_{\nu}$ ($\mathrm{erg~cm^{-2}~s^{-1}}$), with observer time in days. The parameters used can be found in table~\ref{tab:24_run_params}.}
    \label{fig:PR2_plot}
\end{figure*}

Across these top-hat cases, the temporal slopes of the TeV emission are in most cases not impacted by the KN injection and are primarily driven by the dynamics of the GRB jet (the exception being $\epsilon_{e} = 10^{-4}$, where the KN is predominates so strongly that the KN evolution fully controls the TeV emission). The timing of the KN's impact depends on how long it takes for the KN to dominate the IC up-scattering, the initial emission time set in the simulation and how long it takes for said emission to reach the back of the shell.

\section{The role of jet structure and orientation}
\label{sec:Jet}

We now consider the effect of KN seed photons on the VHE emission when we vary the jet structure and observer angle. We fix the dynamical, microphysical and KN parameters to those given in table~\ref{tab:24_run_params}, and then perform $12$ runs, consisting of the $3$ jet structures and $4$ observer angles at $0,~0.5,~1$ and $1.5$ times the half-opening angle of the jet ($\theta_{w}$). For all angles other than on-axis, the observer is off-axis with respect to the core opening angle of the jet, and for the top-hat jet, $\theta_{w}$ is set equal to $\theta_{c}$ such that the observer is far off-axis beyond $\theta_{c}$.

In figure~\ref{fig:PR2_plot}, we show the $100$ GeV light curves, with the three plots showing the results from the top-hat (left), Gaussian (middle) and power-law (right) jet structures. The line-style indicates the observer angle with respect to the jet half-opening angle, $\theta_{w}$. Results with and without injected KN seed photons are shown together.

As we showed above in the top-hat cases, the parameters we use allow the KN seed photons to have a moderate impact on-axis, which we see replicated in Fig.~\ref{fig:PR2_plot}. Because $E_{0}$ is fixed for all jet structures, the on-axis results for the $3$ jets are similar, with an initial boost of around an order of magnitude in the KN light curves. We also see in the Gaussian and power-law on-axis light curves that the KN seed photons cause the VHE flux to decrease with a shallower slope, due to the KN emission predominating over the forward shock emission towards the jet edge as $E(\theta)$ decreases. As a result, the no-KN emission will steepen faster as it follows the drop-off in emission expected from jet deceleration and the full structure of the jet coming into view, while with the KN seed photons, the VHE emission will be primarily dictated by the evolution of luminosity and temperature of the KN.

Due to the weaker synchrotron emission outside the jet core, the KN seed photons can predominate over the VHE emission at earlier times if viewed off-axis from the core. We see in the Gaussian and power-law jets over an order of magnitude difference when viewing outside the core at $0.5 ~\theta_{w}$. This effect is more extreme if viewing from the jet edge ($1~\theta_{w}$), where at $\sim80$ minutes the Gaussian light curve is around $5$ orders of magnitude higher and the power-law light curve is just over $2$ orders of magnitude higher if KN seed photons are included. By $5$ days both cases converge to around $1-2$ orders of magnitude. We therefore see the same effect as found in the $\epsilon_{e}$ and $E_{\rm iso}$ runs; the Gaussian jet, due to the lower $E(\theta)$ and $\gamma$ at the jet edge, has weaker synchrotron emission and has its peak VHE flux more strongly boosted than with the power-law jet, which has stronger synchrotron emission. However, the power-law KN impacted emission is overall stronger as it benefits from the increased number of electrons to up-scatter with, due to the higher $\gamma$ at the jet edge.

Going fully off-axis, or outside $\theta_{w}$, we observe similar boosting as before, though with an additional temporal effect. Due to the far off-axis emission taking longer to reach the observer (see eq.~\ref{eq:t_obs}), there will be an offset between the emission from the jet edge which is impacted by KN seed photons, and the isotropic KN emission reaching the observer. As a result, for a given observer time, we see the effects of the KN seed photons from an earlier lab time at a later observer time. This include times that precede our initial injection of seed photons, which we set to $30$ minutes, leading to an unphysical sharp jump in the KN affected light curves, most apparent in Fig.~\ref{fig:PR2_plot} with the $0.5~\theta_{w}$ top-hat and $1.5~\theta_{w}$ light curves. We exclude any emission before this time in our analysis, as this timescale will be controlled by the rising light curve as the KN becomes optically thin, which is neglected in our model. However, we retain this affected data in the plots above as a visual reference, shown with thin reduced-opacity lines. The full impact on the early far off-axis cannot be modelled without including this rising KN light curve. The Gaussian light curve is less affected by this due to the lower $\gamma$ and thus reduced relativistic beaming.

Once the KN seed photons are injected into the Gaussian jet, we see a shallow, rising light curve, followed by a slight steepening from around $1.5$ days onwards as the tip of the jet, whose VHE output is also dominated by EC emission, comes into view. The power-law jet shows a steeper rising light curve (from $\sim 0.7$ to $2$ days) due to the steeper increase in $\gamma$ as we move to lower latitudes. For all three respective jet structures, the observer will ultimately sees the same underlying jet structure as the outflow decelerates and allows the full jet to become visible, leading to the convergence of the light curves at all angles.

In the top-hat case, we are far off-axis from $0.5~\theta_{w}$ (as $\theta_{c}$ is set as the boundary of the top-hat jet), and so see a similar unphysical jump in this case. Beyond this jump the KN affected light curve follows the same slope as the no-KN slope, though raised by around $1$ order of magnitude as with the on-axis light curves. Very far off-axis, we see that the light curves are either too weak or minimally impacted by the KN emission to be of interest.

\section{Comparison with GRB 170817A}
\label{sec:RealGRBs}
\begin{table}
\caption{Parameters used for GRB 170817A. Afterglow parameters are from \citet{MyPaper}, with $\gamma_0$ set to $200$ in this paper. The kilonova was set by using the broken-power law fitted by \citet{2018MNRAS.481.3423W}, and which is given by eq.~\ref{eq:BPL_L}}
\label{tab:real_GRBs}
\centering
\begin{tabular}{lccc}
\hline
Parameter & Value & Description \\
\hline
$E_{0}$ & $2.16\times10^{53}$ erg & Jet isotropic equivalent energy \\
$\theta_{c}$ & $3.21$\textdegree ~ ($0.056$ rad) & Jet core angle \\
$\theta_{w}$ & $15.47$\textdegree ~ ($0.27$ rad) & Wing truncation angle \\
&  & (power-law/Gaussian only) \\
$n_{\rm ext}$ & $3.6 \times 10^{-3} \rm cm^{-3}$ & External number density \\
$\gamma_{0}$ & $200$ & Initial coasting Lorentz factor \\
$R_{\rm fire}$ & $10^{8}$ cm & Fireball radius \\
z & $0.00973$ & Cosmological redshift \\
$d_{L}$ & $ 1.34 \times 10^{26}~\rm cm$ & Luminosity distance \\
p & $2.13$ & Spectral slope \\
$\epsilon_{e}$ & $1.71 \times 10^{-3}$ & Electron energy fraction \\
$\epsilon_{B}$ & $5.75 \times 10^{-4}$ & B-field energy fraction \\
$L_{0}$ & Eq.~\ref{eq:BPL_L} & Normalised luminosity \\
$T_{0}$ & $5.08\times 10^{4} \rm ~K$ & Normalised temperature \\
$t_{0}$ & 1800 s & Reference time for $L_{0},T_{0}$ \\
$\alpha_{L}$ & Eq.~\ref{eq:BPL_L} & Luminosity power-law index \\
$\alpha_{T}$ & $0.54$ & Temperature power-law index \\
$\chi_{N}$ & $1$ & Fraction of accelerated electrons \\
$\eta$ & $1$ & $\gamma_{\rm max}$ scale factor \\
\hline
\end{tabular}
\end{table}

\begin{figure}
    \includegraphics[scale=0.25, trim={1.8cm 0.7cm 0cm 0}]{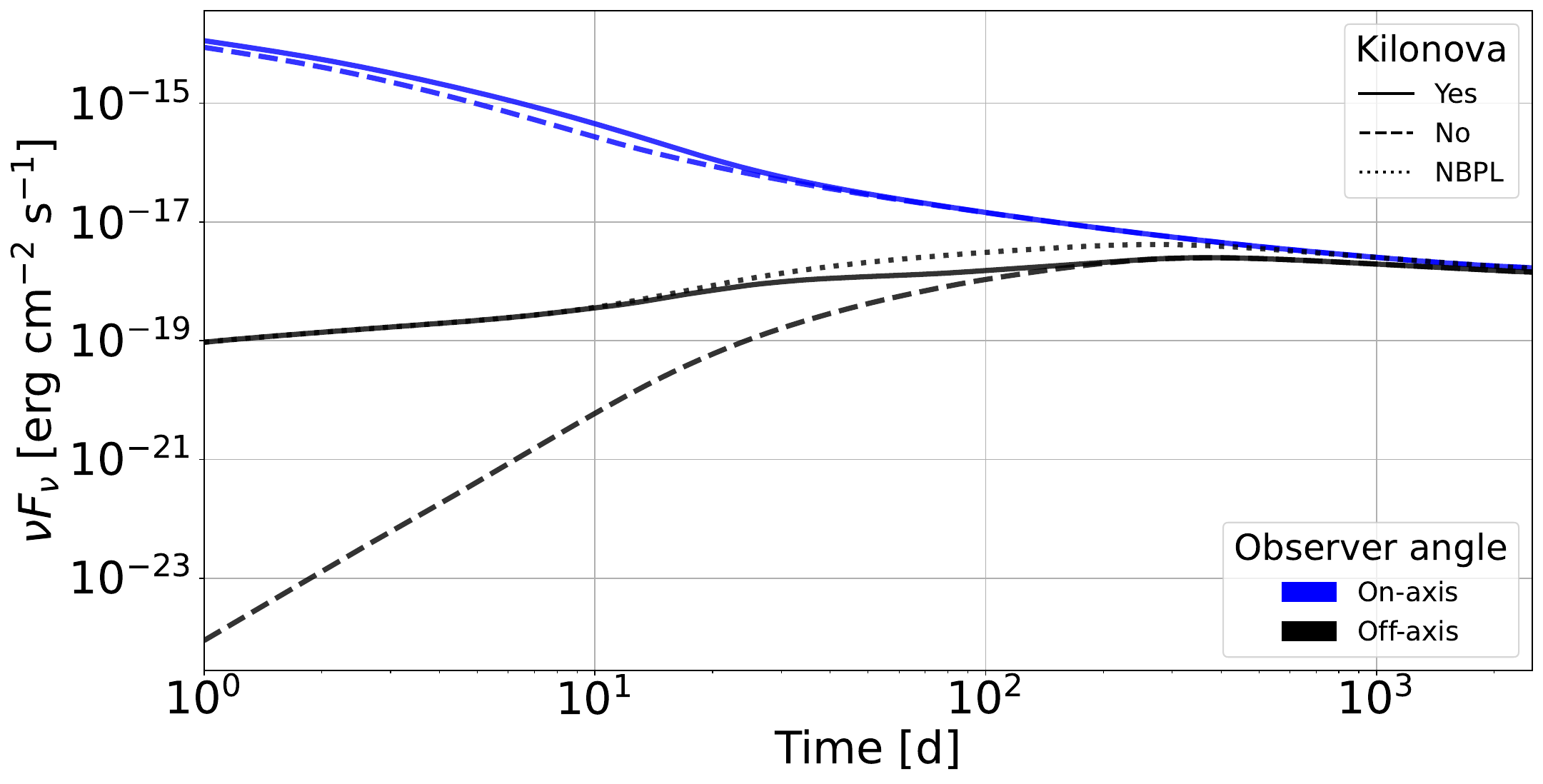}
    \caption{1 TeV light curves for the 170817A runs. We compare the off-axis and on-axis runs with (solid line) and without (dashed line) seed photons injected from the kilonova component. Off-axis with no break in the power-law (NBPL, dotted) is also provided for context. We observe a small boost in the on-axis scenario, but if a kilonova is applied to the off-axis case (ie. the best fitting observer angle), then the TeV flux is significantly boosted until 100 days.}
    \label{fig:LC_170817A}
\end{figure}

We applied this comparison of synchrotron-only vs. KN seed photon boosted emission to the case of GRB 170817A, which our off-axis results from the previous section indicate may have had its VHE emission boosted by its associated KN, AT 2017gfo \citep{Abbott_2017_2}. In the case of 170817A, it was found by \citet{2018MNRAS.481.3423W} that the bolometric luminosity of the KN was best fitted using a broken power-law, which can be given by
\begin{equation}
L(t) = \left\{ \begin{array}{ll} 
  A\left(t/t_{0}\right)^{-\alpha_{L_{1}}} & \text{if } t < t_{b}, \\
  A\left(t_{b}/t_{0}\right)^{\alpha_{L_{2}}-\alpha_{L_{1}}}\left(t/t_{0}\right)^{-\alpha_{L_{2}}} & \text{if } t \geq t_{b},
  \end{array}
  \right.
\label{eq:BPL_L}
\end{equation}
where $A = 6.1\times 10^{41} \times 48^{\alpha_{L_{1}}}$ is the normalised bolometric luminosity at $t_{0} = 1800$ seconds ($30$ minutes), $t_{b}$ is the break time (in seconds) at $6.2$ days, $\alpha_{L_{1}} = 0.95$ is the first power-law slope and $\alpha_{L_{2}} = 2.8$ is the second power-law slope. The temperature follows a simple power-law. The afterglow and temperature parameters used in this run are given in table~\ref{tab:real_GRBs}, with the afterglow parameters set according to the rescaled parameters fitted in \citet{MyPaper}. Four runs were performed, with two runs with no KN seed photons and two with seed photons. They were further divided into runs where we observe 170817A on-axis and with the expected off-axis viewing angle. The results from these runs are shown in Fig.~\ref{fig:LC_170817A}.

Looking at the off-axis scenario, as with the previous section, we observe a significant boost to the TeV flux of around $5$ orders of magnitude at $1$ day, dropping to $1-2$ orders of magnitude by $10$ days. The rising light curve shows a shallower slope of $0.5$ as opposed to the no-KN case with $3.8$. The slope in the KN case slightly steepens around $10-20$ days before shallowing to $0.3$, causing it to converge with the no-KN case from $\sim 100$ days onwards. This final shallowing is due to the power-law break in the luminosity, as if we neglect to include that break (see the dotted line), this shallowing of the slope is not observed and the off-axis flux does not converge until around $1000$ days onward.

For the on-axis case, we observe a small increase in TeV emission, maximally varying from the no-KN case most at $5-10$ days, though with minimal impact to the peak flux. None of the TeV light curves would be visible, even by the Cherenkov Telescope Array Observatory (CTAO)\footnote{See https://www.ctao.org/for-scientists/performance/}, which has a lower sensitivity limit of $\sim10^{-13}~\rm erg~cm^{-2}~s^{-1}$ at energies of $1$ TeV after $50$ hours of observation.

The KN seed photons do not have any impact on the time of convergence between the on-axis and off-axis measurements, which converge as before at $\sim 1000$ days.

\section{Discussion}
\label{sec:Discuss}

\begin{figure*}
    \centering
    \includegraphics[width=0.9\linewidth]{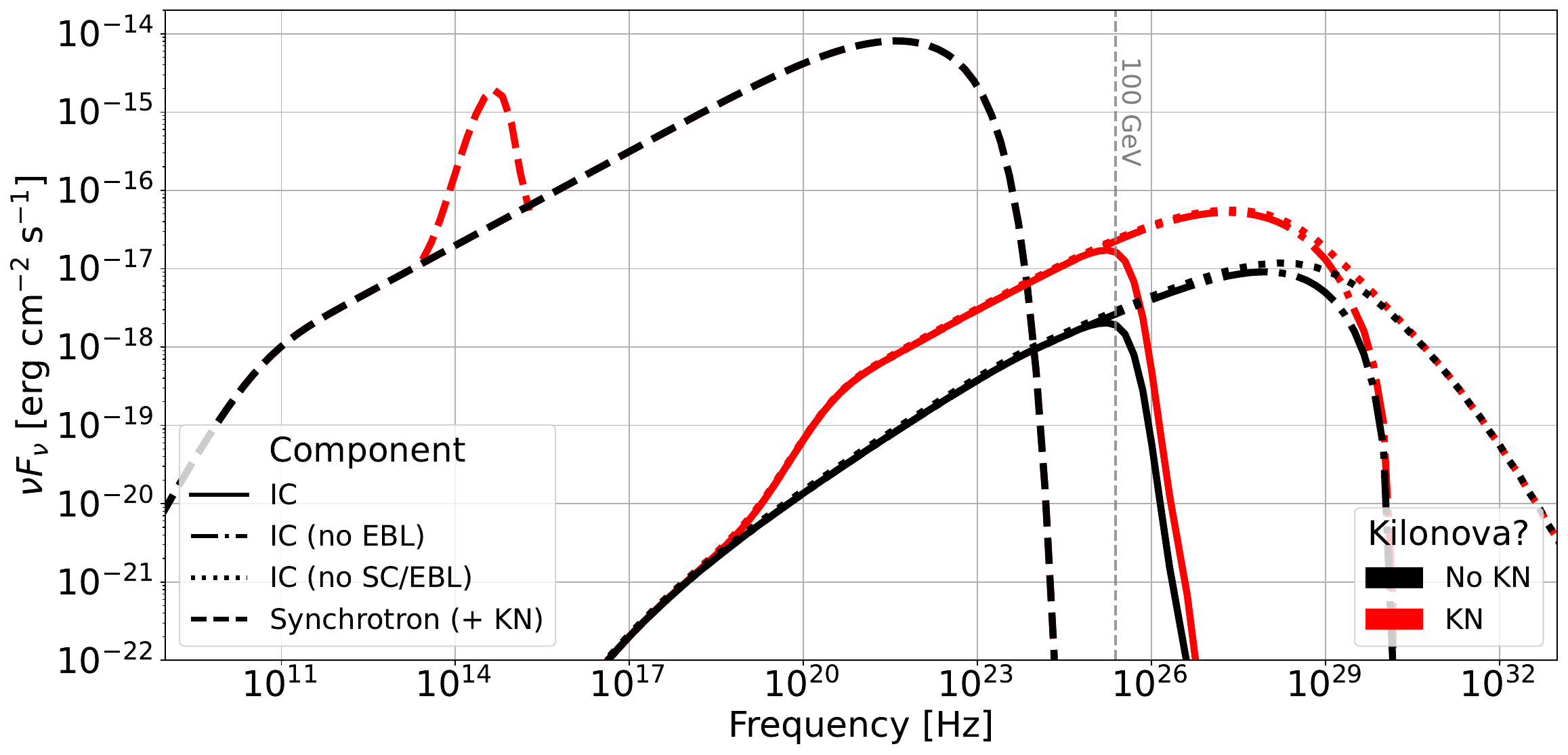}
    \caption{Spectra from the table~\ref{tab:24_run_params} scenario at $4\times10^{4}\rm~s$. The synchrotron (dashed, with cut-off), IC (solid), IC with no EBL attenuation (dot-dash) and IC with no EBL attenuation or synchrotron cut-off (dotted) emissions are shown separately, and a grey dashed line marks $100$ GeV. Emission with and without injected KN seed photons are shown together. All results are given in units of $\nu F_{\nu}$ ($\mathrm{erg~cm^{-2}~s^{-1}}$), with frequency in Hz.}
    \label{fig:SYNIC}
\end{figure*}

Overall, we see that the VHE emission can be impacted by the KN seed photons, with the strength of the impact determined by the ability of the KN flux to exceed the afterglow emission in the optical/UV regime. However, the broadband emission in the above cases are all very low, due to the parameters used in this work. The VHE emission is further suppressed by the combined effect of the EBL attenuation and the Klein-Nishina effect on the IC up-scattering. In the above parameter and jet structure cases, the primary effect of the KN seed photons is to mitigate the loss of VHE emission as we move to regions of the parameter space where the VHE emission is lower due to the rightward movement of $\nu_{c}$, which is $\propto \epsilon_{B}^{-3/2}n_{\rm ext}^{-1}E_{_{\rm iso}}^{-1/2}$ \citep{Granot_2002} and weaker IC up-scattering from the afterglow parameters (such as low $\epsilon_{e}$ or $n_{\rm ext}$).

\subsection{Impact on the IC emission}
\label{sec:Impact}
The exact impact of the KN seed photons on the spectrum may be hard to see when all effects (such as EBL attenuation, summation of synchrotron and IC components) are taken together. In Fig.~\ref{fig:SYNIC}, we show the separate components of the synchrotron, IC and KN emission in the spectrum. We also include the cases where EBL attenuation is neglected as this allows the IC component in the electron-photon gas to be more clearly seen. The scenario used is the typical sGRB parameters from table~\ref{tab:24_run_params}.

Without the synchrotron emission predominating over the IC component, we can now see the rise in the IC component due to the KN emission being included, starting at $\sim10^{19}~\rm Hz$ and continuing elevated from $\sim 10^{21}\rm~Hz$. This corresponds to IC up-scattering of the peak KN seed photons by electrons at $\gamma_{m}$; since in the Thomson limit $\nu_{\rm obs,~IC} \sim \gamma_{m}^{2}\nu_{\rm BB}$, and from our model, $T_{\rm eff} = 9.65\times10^{3} ~\textrm{K} \sim h\nu_{\rm BB}/(2.82k_{B})$ and $\gamma_{m}\sim227$, the seed photons will be mostly up-scattered to $\nu_{\rm obs,~IC} \sim 2.91\times 10^{19}\rm ~Hz$. The initial rise in IC emission with KN seed photons around $10^{19-20}$ Hz will be controlled by the broadness of the thermal emission spectrum, and the rising slope thereafter will be due to the seed photons scattering off the non-thermal ensemble of electrons up to $\nu_{c}$ and the synchrotron cut-off.

It should be stressed again that the primary driver of the peak \textit{observed} IC emission in both KN and no-KN scenarios is the EBL attenuation; both see their IC emission peak at the same energy of $\sim 100 \rm~GeV$, before they both fall off at the same rate. Without EBL attenuation, the peak emission for the seed photons is controlled by Klein-Nishina suppression. We can show this by invoking the work done by \citet{NakarKN} (see also \citealt{McCarthy_2024}); unlike the Thomson limit, the peak will be constrained, not by the cooling break, but by the maximum electron energy $\hat{\gamma}_{\rm BB}$ available to up-scatter with before Klein-Nishina suppression becomes important. This is given by
\begin{equation}
\label{eq:gamma_hat}
    \hat{\gamma}_{\rm BB} = \frac{m_{e}c^{2}\gamma}{h\nu_{\rm BB}}. 
\end{equation}
We can use this to approximate what the peak IC emission should be; from our shell model, at $4\times10^{4}\rm~s$, $\gamma \sim 8.51$, and from $T_{\rm eff}$, $\nu_{\rm BB} = 5.67\times 10^{14}\rm~Hz$. This gives us $\hat{\gamma}_{\rm BB} \sim 1.85\times10^{6}$. Applying this critical energy to the IC up-scattering of $\nu_{\rm BB}$ then gives us $\nu_{\rm obs,~IC_{\rm max}}\sim \hat{\gamma}_{\rm BB}^{2}\nu_{\rm BB} = 1.93\times 10^{27}\rm~Hz$, which is comparable to the simulation result, where we see the IC emission peak at $2.15 \times 10^{27}~\rm Hz$. The peak emission will then evolve over time, with $\nu_{\rm obs,~IC_{\rm max}} \propto t_{\rm obs}^{-2.5}$.

If SSC is the only source of photons, then the resulting IC emission peaks lower in flux, and around an order of magnitude higher in frequency. This also arises from Klein-Nishina suppression; from \citet{NakarKN}, when in synchrotron slow cooling, the peak IC emission will be $\nu_{\rm obs,~IC_{\rm max}} \sim 2\nu_{c}\gamma_{c}\hat{\gamma}_{c}$. From theory, $\gamma_{c} = 6\pi \gamma m_{e}c / (\sigma_{T}B^{2}t_{\rm lab}) = 8.41\times 10^{7}$ ($B = 1.7\times10^-3$ and $t_{\rm lab} = 2.71\times10^{7}\rm~s$ are found from our shell model; this equation does not include IC cooling, though in this case the correction is minor), which will give us the cooling break frequency $\nu_{c} \sim 2\gamma(1 + z)^{-1} \times 3q_{e}\gamma_{c}^{2}B/(4\pi m_{e}c) = 2.62 \times10^{20}~\rm Hz$. The final term is of the same form as eq.~\ref{eq:gamma_hat}, and is $\hat{\gamma}_{c} = m_{e}c^{2}\gamma/(h(1+z)\nu_{c}) = 2.35$. Therefore, we get $\nu_{\rm obs,~IC_{\rm max}} \sim 1.08 \times 10^{29}\rm ~Hz$. By comparison, the kinetic model predicts a lower value of $1.47 \times 10^{28}~\rm Hz$ as seen in Fig.~\ref{fig:SYNIC}. Note that the expected Thomson up-scattered peak $\nu_{c,~\rm IC_{Thom}} \sim \gamma_{c}^{2}\nu_{c} = 1.86\times10^{36}~\rm Hz$ lies well beyond the range covered by Fig.~\ref{fig:SYNIC}.

In summary, EBL attenuation is the strongest constraining factor for peak emission in both KN and no-KN scenarios. If EBL attenuation is neglected, then Klein-Nishina suppression becomes the key constraining effect on the peak IC emission for both KN and SSC photons, albeit only at frequencies above $\sim 10^{27}$ Hz in our example. If both EBL attenuation and synchrotron cut-off are neglected, then the emission will not have an exponential cut-off and both will have a shallower decrease in IC emission in the Klein-Nishina limit, with the KN seed photon emission converging with SSC emission at $10^{30}\rm~Hz$. We emphasise that the physically expected synchrotron cut-off (as set by Eq.~\ref{eq:gamma_max}) \citep{Dai_2001} is crucial for the IC emission to stand out against any extrapolated synchrotron emission.

\subsection{Constraining the KN/afterglow parameters}
\label{sec:Constraint}
It is helpful to consider the constraints required for the KN seed photons to have a meaningful impact on the VHE emission. Consider the luminosity of the KN; for it to have an impact, it must predominate over the afterglow emission in the optical/UV regimes, $F_{\nu,~\rm BB} >> F_{\rm Opt/UV}$. An approximation of this can be obtained by comparing the evolution of the peak spectral luminosity as derived in appendix~\ref{App:KN} to the afterglow flux prescriptions provided by \citet{Granot_2002}. Since the peak KN spectral frequency $\nu_{\rm BB,~max} \propto T \propto t^{-\alpha_{T}}$, and in the (slow cooling) afterglow, the minimum injection frequency $\nu_{m} \propto t_{\rm obs}^{-3/2}$, the 
self-absorption frequency $\nu_{a} \propto t_{\rm obs}^{-0.6}$ (if $\nu_{m} < \nu_{a}$, $\sim t_{\rm obs}^{-0.7}$) and the cooling break $\nu_{c} \propto t_{\rm obs}^{-1/2}$, we can argue that $\nu_{m,~a} < \nu_{\rm BB,~max} < \nu_{c}$ for all times of interest if $\alpha_{T} \lesssim 0.5$; this holds true in all the cases considered above. This means we can limit our analysis to region G from \citet{Granot_2002} for all times of interest.

The flux from the KN is given by
\begin{equation}
F_{\nu, {\rm BB}} = \frac{L_\nu}{4 \pi d_L^2} = \frac{L B_\nu}{4 \pi d_L^2 \sigma T^4},
\end{equation}
which, for the peak flux, can be recast into the form
\begin{eqnarray}
F_{\nu,BB} (\nu \approx \frac{k_B T}{ 2.82 h}) &= &5.6 \times 10^{-31} \left( \frac{d_L}{10^{28} \textrm{ cm}} \right)^{-2} \\
& &\left( \frac{L}{1.6 \times 10^{42} \textrm{ erg s}^{-1}} \right) \nonumber \\ &&\left( \frac{T}{7.5 \times 10^3 \textrm{ K}} \right)^{-1} \textrm{ erg cm}^{-2} \textrm{ s}^{-1} \textrm{ Hz}^{-1}, \nonumber
\end{eqnarray}
where the reference values used have been taken from our standard sGRB scenario from table~\ref{tab:24_run_params} at $4\times 10^{4}\rm ~s$. From region G in \citet{Granot_2002}, we obtain
\begin{eqnarray}
F_G &= &2.7 \times 10^{-32} \left( \frac{\epsilon_e}{10^{-1}}\right)^{p-1} \left( \frac{\epsilon_B}{10^{-4}}\right)^{\frac{1+p}{4}} \nonumber \\
& & \left( \frac{E_0}{10^{51} \textrm{ erg}} \right)^{\frac{3+p}{4}} \left( \frac{n_{\rm ext}}{3 \times 10^{-3} \textrm{ cm}^{-3}} \right)^{\frac{1}{2}} \left( \frac{t}{4 \times 10^4 \textrm{ s}} \right)^{\frac{3(1-p)}{4}} \nonumber \\
& & \left( \frac{d_L}{10^{28} \textrm{ cm}} \right)^{-2} \left( \frac{T}{7.5 \times 10^3 \textrm{ K}} \right)^{\frac{1-p}{2}},
\end{eqnarray}
using $p = 2.2$ to set the numerical pre-factor. In both KN and afterglow flux, we neglect the cosmological redshift factor.

Therefore, for the KN to have an impact on the VHE emission, 
\begin{align}
\left( \frac{L}{1.6 \times 10^{42} \textrm{ erg s}^{-1}} \right) \left( \frac{T}{7.5 \times 10^3 \textrm{ K}} \right)^{\frac{p-3}{2}} > 4.8 \times 10^{-2} \nonumber \\ 
\times \left( \frac{\epsilon_e}{10^{-1}}\right)^{p-1} \left( \frac{\epsilon_B}{10^{-4}}\right)^{\frac{1+p}{4}} \left( \frac{E_0}{10^{51} \textrm{ erg}} \right)^{\frac{3+p}{4}} \nonumber \\ 
\times \left( \frac{n_{\rm ext}}{3 \times 10^{-3} \textrm{ cm}^{-3}} \right)^{\frac{1}{2}} \left( \frac{t}{4 \times 10^4 \textrm{ s}} \right)^{\frac{3(1-p)}{4}}.
\label{eq:KN_G_comparison}
\end{align}

For the reference values chosen from this work, the KN will exceed the afterglow by a factor of $\sim 20$. As seen in Fig.~\ref{fig:Series_plot}, this then generates a significant boost in the VHE emission. At high $n_{\rm ext}$, $E_{\rm iso}$ and $\epsilon_{B}$, the afterglow emission is sufficiently boosted that this cannot occur, even at later times. This can be mitigated by a lower temperature or higher luminosity.

The luminosity at early times will be controlled by the lanthanide-poor ("blue") KN emission, producing a brighter, earlier emission with a higher temperature. At later times on the scale of days-weeks, the lanthanide-rich ("red") KN will produce a less luminous component with a lower temperature. The combination of both components has been used to explain the evolution of AT2017gfo (as outlined by \citet{MetzgerLR} in his review; see also for example, \citealt{2017Natur.551...80K, 10.1093/pasj/psx121, doi:10.1126/science.aap9455}). A blue component is preferential for VHE emission, as eq.~\ref{eq:KN_G_comparison} shows that a higher luminosity will overcome the higher temperature, due to the luminosities stronger linear dependence (though if off-axis, as in GRB 170817A, the impact of the red component can be seen more clearly). Additionally, beyond a few days, the afterglow will have decayed to the point where the photon/electron gas will not support any detectable up-scattering even if the red component is considered -- the latest VHE emission detected from a GRB afterglow was around $52$ hours from the long GRB 190829A \citep{doi:10.1126/science.abe8560}. No short GRBs have yet been detected with a VHE emission beyond a day.

One possible solution to boost the bolometric luminosity is by considering magnetar-driven KN\footnote{Also known as a merger-nova.} \citep{Yu_2013}, which could have peak bolometric luminosities of $10^{43}-10^{44} \rm ~ergs^{-1}$ at times up to $\sim 1$ day later \citep{10.1093/mnras/stu247, 10.1093/mnras/stac2609}, allowing for the KN seed photons to exceed late time afterglow emission and re-brighten or sustain the VHE emission for much longer. In this case, the rising light curve of the bolometric luminosity becomes important and would need to be modelled.

Currently, no VHE emission from GRBs has been associated with KN. It was considered in the case of 211211A by \citet{2022Natur.612..236M}, but was dismissed due to the afterglow being predominant and its VHE emission being negligible in the expected range for KN seed photons. Instead, it was argued that the KN seed photons were being up-scattered in a low power jet behind the blast-wave which was closer to the KN. Another case with excess VHE emission was from 160821B, though it was argued by \citet{Zhang_160821B} that the seed photons in this case arose from extended and plateau emission. Indeed, as mentioned in the introduction, there are several mechanisms that can generate seed photons for the forward shock, which must also be considered when accounting for excess VHE emission.

\subsection{Limitations of this work}
We note that there are a few limits to our analysis that we lay out here:
\begin{itemize}
    \item \textbf{Power-law KN:} we have limited our analysis to the the power-law phase of the KN emission, as found for GRBs 211211A \citep{Troja211211A} and 1701817A \citep{2018MNRAS.481.3423W}. In reality, there will be an initial rising phase of emission which will be heavily dependent on effects like neutron heating \citep{MetzgerKNprecursor}, the opacity of the KN during this phase \citep{10.1093/mnras/staa1576, Banerjee_2020} or shock re-heating \citep{MetzgerLR}. A better understanding of the very early stages of the KN must therefore model these effects, which we leave to future work.
    \item \textbf{Homogeneity of shell:} while we have made use of a multi-zone model in this work that accounts for lateral structure, each zone is itself still treated as isotropic and homogeneous. One consequence of this simplification is that any seed photons injected into the back of the shell (from the KN) will be treated as simultaneously present across the shell, which can lead to (for example in the case of the $\epsilon_{e}$ light curves Fig.~\ref{fig:Series_plot}) to emission that appears earlier than the $30$ minute mark, as the shell initially keeps up with escaping photons. However, this is an issue that primarily affects the initial stages of this injection, which as we have just mentioned are also impacted by the early-time evolution of the KN.
\end{itemize}

\section{Conclusions}
\label{sec:Conclude}
In this work, we have applied a basic KN model to our GRB afterglow kinetic code to explore the effects of KN seed photons on the VHE emission of the afterglow. We obtained the following conclusions:

\begin{itemize}
    \item \textbf{Afterglow parameters} -- In the case of typical sGRBs, we saw a significant boost to the VHE emission, around $1$ order of magnitude, when seed photons from the KN were included. However, the boost in VHE emission achieved in this manner was not beyond the level producible by SSC using parameter values that led to a brighter synchrotron afterglow. For $\epsilon_{B}$ and $n_{\rm ext}$, this caused the TeV emission across all tests to be comparable with seed photon injection, while in the $\epsilon_{e}$ and $E_{\rm iso}$ runs, this effect arrested the fall in VHE emission when the latter two parameters were lowered. Conversely, where the afterglow flux was strongest in parameter space, the KN seed photons had no impact. We showed through eq.~\ref{eq:KN_G_comparison} how the KN and afterglow parameters could be used to determine if the KN seed photons would be impactful.
    \item \textbf{Jet structure and orientation} -- When structure was introduced to the GRB jet, the lower $E(\theta)$ at higher jet angles caused these regions to be substantially affected by the KN seed photons. This led to a shallower VHE emission when observing the Gaussian and power-law jets on-axis, than otherwise expected for SSC. This effect was further boosted off-axis, where the jet edge controls the initially observed emission. Combined with the earlier emission arriving later when the observer is far off-axis, this can create a substantially higher VHE emission than otherwise expected if KN emission was not accounted for.
    \item \textbf{GRB 170817A} -- The inclusion of the KN seed photons for GRB 170817A significantly boosted the $1$ TeV light curve, with a $5$ order of magnitude increase at $1$ day, and a shallower light curve than the no-KN case. The broken power-law, which is thought to be due to the red component becoming dominant in the KN \citep{MetzgerLR}, led to a further shallowing beyond $30$ days, with a convergence of the KN and no-KN cases beyond $100$ days. The on-axis emission was minimally impacted.
\end{itemize}

The overall effect of KN seed photons we have seen throughout is to help weaker afterglows in up-scattering photons by providing an additional supply of photons to up-scatter with. However, this does not necessarily overcome the limits imposed by an electron population which is smaller and less energetic. While in this work none of the cases we have discussed would impact the VHE emission observable by current instruments, it remains possible, particularly when considering effects such as magnetar driven KN, that VHE emission could be impacted. In either case, if eq.~\ref{eq:KN_G_comparison} holds true, the effect of KN emission on the VHE afterglow should be considered alongside other mechanisms such as late time prompt emission and the reverse shock.

We therefore conclude that KN seed photons are a credible and promising source of seed photons for boosting VHE emission, especially in the case of off-axis structured jets. However, standard sGRBs are not suitable scenarios for this effect to be observed, and atypical sGRBs (such as an on-axis GRB 170817A) along with magnetar-driven KN are more likely to provide the conditions necessary for the impact of KN seed photons to be seen.

\section*{Acknowledgements}

The authors would like to thank the referee for their constructive feedback and suggestions for this paper. JPH acknowledges funding from the Science and Technology Facilities Council (STFC) through grant ST/W507301/1. HJvE further acknowledges support by the European Union Horizon 2020 programme under the AHEAD2020 project (grant agreement number 871158).

\section*{Data Availability}
No new data was generated or analysed in support of this research.



\bibliographystyle{mnras}
\bibliography{refs} 




\appendix

\section{Derivation of kilonova photon injection}
\label{App:KN}
We can approximate the emission of the kilonova as a blackbody, with a bolometric luminosity $L$ and effective temperature $T$. Standard textbook theory (see e.g. \citealt{RybLi}) tells us that the spectral intensity for a blackbody $B_{\nu}$ is given by
\begin{equation}
    \label{eq:BB_Inst}  
    I_\nu = B_\nu = \frac{2h\nu^3}{c^2} \frac{1}{\exp \left( \frac{h \nu}{k_B T} \right) - 1},
\end{equation}
where $\nu$ is the frequency in the burster/lab frame, $h = 6.626 \times 10^{-27}~ \mathrm{erg~s}$ is Planck's constant and $k_{B} = 1.38 \times 10^{-16}~ \mathrm{erg~K^{-1}}$ is Boltzmann's constant.

The spectral flux is then the integral over solid angle of the spectral intensity, $F_{\nu} = \pi B_{\nu}$.

Since the spectral luminosity $L_{\nu} = 4 \pi R^{2} F_{\nu}$, we can obtain the bolometric luminosity:
\begin{align}
    L_{\nu} &= 4\pi^{2} R^{2}B_{\nu}, \label{eq:Lnu} \\ 
    L &= 4 \pi^2 R^2 \int B_\nu d\nu = 4 \pi R^2 \sigma T^4, \label{eq:Lall}
\end{align}
where $\sigma = 5.67 \times 10^{-5}~\mathrm{erg~cm^{-2}~s^{-1}~K^{-4}}$ is the Stefan-Boltzmann constant.

Now consider a GRB afterglow shell moving outwards with a fluid Lorentz factor $\gamma$ at a distance $R$ from the burster. The energy per unit time per unit frequency being received by the shell is then given by
\begin{equation}
\label{eq:dEdndt_Lnu}
    \frac{dE}{d \nu dt} = L_\nu \left( t \right) \left( 1 - \beta_{\textrm{back}} \right) \frac{\Omega_{\textrm{back}}}{4 \pi},
\end{equation}
where $t = t - \frac{R_{\rm back}}{c}$ is the emission time of the kilonova (as distinct from the arrival time $t$ when the emission arrives at the back of the shell at distance $R_{\rm back}$) and $\Omega_{\rm back}$ is the solid angle of the receiving shell. Since we are considering isotropic shells, $\Omega_{\rm back}$ may be set to $4\pi$. $\beta_{\rm back}$ is the relative velocity of the back of the afterglow shell,
\begin{equation}
    \beta_{\rm back} = \frac{1}{c}\left( \skew{3}\dot{R}\left(1 - \frac{1}{12\gamma^{2}}\right) + \frac{\dot{\Gamma}R}{6\gamma^{3}}\right),
\end{equation}
where $\Gamma \sim \sqrt{2}\gamma^{2}$ is the forward shock Lorentz factor, and we have accounted for the expansion of the shell width in this equation. The inclusion of the $(1 - \beta_{\rm back})$ term is to account for the fraction of light that fails to catch up with the relativistic shell at a given arrival time.

We can rewrite the above equation in terms of the bolometric luminosity by combining eqs. \ref{eq:Lnu} \& \ref{eq:Lall} with \ref{eq:dEdndt_Lnu} to get
\begin{equation}
    \frac{dE}{d \nu dt} = \frac{\pi B_{\nu} \left(t\right)L\left(t\right)}{\sigma T\left(t\right)^{4}}\left(1 - \beta_{\rm back}\right).
\end{equation}

We must translate this into the frame of reference of the expanding shell. Since the emitter is front-back symmetric in the frame of the emitter, we can state that the total emitted power is Lorentz invariant (\citealt{RybLi}). Accounting for the Doppler shift in frequency $d\nu = \gamma \left( 1 + \beta \right) d \nu^{\prime}$, we thus get
\begin{equation}
    \frac{dE^{\prime}}{dt\prime d \nu\prime} = \gamma \left( 1 + \beta \right) \frac{\pi B_\nu \left( t \right) L \left( t \right)}{\sigma T\left( t\right)^4} \left( 1 - \beta_{\textrm{back}}\right).
\end{equation}

The energy per unit frequency can be related to the number of particles as
\begin{equation}
    \frac{dE^{\prime}}{d\nu^{\prime}} = h\nu^{\prime}\frac{dN^{\prime}}{d\nu^{'}} = h\nu^{\prime}V^{\prime}\frac{dn^{\prime}}{d\nu^{\prime}},
\end{equation}
where we have also assumed that the volume of the shell at an instantaneous time is $V^{\prime}$, giving us a co-moving number density $n^{\prime}$ (here $\prime$ is used to indicate fluid frame). Combining this with the definition for $B_{\nu}$ in eq.~\ref{eq:BB_Inst}, as well as noting that the dimensionless photon energy $\epsilon^{\prime} = h\nu^{\prime} / m_{e}c^{2}$, we finally get
\begin{align}
    \frac{dn^{\prime}}{dt^{\prime}d\epsilon^{\prime}} & =     \left(\frac{m_{e}c^{2}}{h}\right)^{3}
    \frac{2{\epsilon^{\prime}}^{2}}{V^{\prime}c^{2}}
    \frac{\gamma^{4}\left(1 + \beta\right)^{4}}{\exp(\frac{\gamma(1 + \beta)m_{e}c^{2}\epsilon^{\prime}}{k_{B}T\left(t\right)}) - 1} \nonumber \\
    &  \times 
    \frac{\pi L\left(t\right)}{\sigma T\left(t\right)^{4}}(1 - \beta_{\rm back}).
\end{align}

\bsp	
\label{lastpage}
\end{document}